\documentclass[aps,prl,showpacs,floatfix,twocolumn,byrevtex,superscriptaddress]{revtex4-1}
\usepackage{amsmath}
\usepackage{amssymb}
\usepackage{amstext}
\usepackage{amsopn}
\usepackage{amsfonts}
\usepackage{amsxtra}
\usepackage[english]{babel}
\usepackage{graphicx}
\usepackage{float}
\usepackage{bm}
\usepackage{multirow}
\usepackage{dcolumn}
\usepackage{color}
\usepackage{hyperref}

\newcommand{\icm}{\ensuremath{\textrm{cm}^{-1}}}% % cm-1

\newcommand{\NNO}{Nd$_{4}$Ni$_{3}$O$_{8}$}
\newcommand{\LNO}{La$_{4}$Ni$_{3}$O$_{8}$}
\newcommand{\PNO}{Pr$_{4}$Ni$_{3}$O$_{8}$}
\newcommand{\RNO}{$R_{4}$Ni$_{3}$O$_{8}$}

\begin{document}

\title{Charge-stripe fluctuations in Nd$_{4}$Ni$_{3}$O$_{8}$ as evidenced by optical spectroscopy}
\author{Jiahao Hao}
\author{Xinwei Fan}
\author{Qing Li}
\author{Xiaoxiang Zhou}
\author{Chengping He}
\author{Yaomin Dai}
\email{ymdai@nju.edu.cn}
\affiliation{National Laboratory of Solid State Microstructures and Department of Physics, Collaborative Innovation Center of Advanced Microstructures, Nanjing University, Nanjing 210093, China}
\author{Bing Xu}
\affiliation{Beijing National Laboratory for Condensed Matter Physics, Institute of Physics, Chinese Academy of Sciences, Beijing 100190, China}
\author{Xiyu Zhu}
\author{Hai-Hu Wen}
\affiliation{National Laboratory of Solid State Microstructures and Department of Physics, Collaborative Innovation Center of Advanced Microstructures, Nanjing University, Nanjing 210093, China}

\date{\today}
%%%%%%%%%%%%%%%%%%%%%%%%%%%%%%%%%%%%
%
% Abstract
%

\begin{abstract}
We present an investigation into the optical properties of Nd$_{4}$Ni$_{3}$O$_{8}$ at different temperatures from 300 down to 5~K over a broad frequency range. The optical conductivity at 5~K is decomposed into IR-active phonons, a far-infrared band $\alpha$, a mid-infrared band $\beta$, and a high-energy absorption edge. By comparing the measured optical conductivity to first-principles calculations and the optical response of other nickelates, we find that Nd$_{4}$Ni$_{3}$O$_{8}$ features evident charge-stripe fluctuations. The $\beta$ band is attributed to electronic transitions between the gapped Ni-$d_{x^2-y^2}$ bands due to fluctuating charge stripes, while the high-frequency absorption edge corresponds to the onset of transitions involving other high-energy bands. Furthermore, an analysis of the temperature-dependent optical spectral weight reveals a $T^{2}$ law, which is likely to originate from strong correlation effects.
\end{abstract}

%  71.55.Ak  Metals, semimetals, and alloys
%  72.15.Eb: Electrical and thermal conduction in crystalline
%  78.20.-e  Optical properties of bulk materials and thin films
%  78.30.-j  Infrared and Raman spectra

%\pacs{78.20.-e, 78.30.-j}

\maketitle

%%%%%%%%%%%%%%%%%%%%%%%%%%%%%%%%%%%%%%%%%%%%%%%%%%%%%%%%%%%%%%%%%%%%%%%%%%%%%%
%
% Introduction
%
Since the discovery of unconventional superconductivity in copper oxide compounds~\cite{Bednorz1986ZPBCM}, tremendous efforts have been devoted to the exploration of cuprate-like materials based on different transition-metal ions in the hope of finding new superconducting systems and gaining insights into the pairing mechanism in cuprates. Due to the proximity of Ni to Cu in the periodic table, nickelates have attracted great attention~\cite{Anisimov1999PRB,Lee2004PRB,Poltavets2010PRL,Zhang2016PNAS,Zhang2017NP,Cheng2012PRL,Pardo2010PRL,Pardo2012PRB,Liu2012JPCM,ApRoberts-Warren2011PRB,Botana2016PRB,Botana2017PRM}. Anisimov et al. have pointed out that nickelate analogs to the superconducting cuprates may be realized only if the Ni$^{+}$ ($3d^{9}$) ions are forced into a planar oxygen coordination, which mimics the parent compounds of the cuprates, and then doped with low-spin Ni$^{2+}$ ($3d^{8}$) holes~\cite{Anisimov1999PRB}. On the other hand, Lee and Pickett~\cite{Lee2004PRB} have compared LaNiO$_{2}$ (Ni$^{+}$, 3$d^{9}$) to CaCuO$_{2}$ (Cu$^{2+}$, 3$d^{9}$), and found very different behavior, such as significantly reduced 2$p$-3$d$ hybridization and the existence of La 5$d$ at the Fermi level in LaNiO$_{2}$, arguing against the analogy between nickelates and cuprates. Recently, superconductivity with a critical temperature ($T_c$) up to 15~K has been reported in hole-doped infinite-layer nickelates Nd$_{1-x}$Sr$_{x}$NiO$_{2}$~\cite{Li2019Nature,Li2020PRL,Zeng2020PRL}, reigniting the debate on whether nickelates represent analogs of cuprates~\cite{Botana2020PRX,Karp2020PRX,Zhang2020PRB,Jiang2020PRL}.

Hole-doped $3d^{9}$ Ni ions in a planar oxygen coordination also exist in the trilayer $R_{4}$Ni$_{3}$O$_{8}$ ($R = $ La, Pr or Nd). These compounds are 1/3 self hole doped with a nominal 3$d^{8.67}$ filling, which should fall into the overdoped Fermi-liquid regime in the phase diagram of hole-doped cuprates~\cite{Botana2017PRM}. From an experimental perspective, while Pr$_{4}$Ni$_{3}$O$_{8}$ seems to be a metal without any phase transition in the temperature range 2--300~K~\cite{Zhang2017NP}, La$_{4}$Ni$_{3}$O$_{8}$ exhibits a semiconductor-insulator transition at about 105~K, resulting in a highly insulating ground state~\cite{Poltavets2010PRL,Cheng2012PRL,Zhang2016PNAS,Zhang2017NP}. The formation of long-range antiferromagnetic order has been detected below the transition at 105~K by La$^{139}$ nuclear magnetic resonance (NMR)~\cite{ApRoberts-Warren2011PRB,Poltavets2010PRL} and neutron diffraction~\cite{Zhang2019PRL}; a recent synchrotron X-ray diffraction study has revealed that the ground state of \LNO\ is a quasi-two-dimensional charge-stripe-ordered insulator~\cite{Zhang2016PNAS}. \NNO\ also shows insulating behavior with no signature of a phase transition from room temperature down to 2~K~\cite{Li2021SCPMA}. Theoretically, the ground state of $R_{4}$Ni$_{3}$O$_{8}$ can be metallic with a low-spin (LS) configuration or insulating with a high-spin (HS) configuration~\cite{Pardo2010PRL,Pardo2012PRB,Liu2012JPCM}, depending on the relative magnitude of Hund's rule coupling ($J_H$) and the crystal field splitting ($\Delta_{\text{cf}}$) between the $d_{x^2-y^2}$ and $d_{3z^2-r^2}$ bands induced by the absence of apical oxygen ions. Moreover, Botana et al. have demonstrated that a low-spin charge stripe (LS-CS) insulating ground state can also be obtained from a combination of charge-order-related structural distortions and magnetic order~\cite{Botana2016PRB}.

To date, questions regarding the ground state of the trilayer $R_{4}$Ni$_{3}$O$_{8}$ compounds, and to what extent they are similar to the cuprates are still open to debate~\cite{Poltavets2010PRL,ApRoberts-Warren2011PRB,Cheng2012PRL,Zhang2016PNAS,Zhang2017NP,Zhang2019PRL,Lin2020arXiv}. We notice that the theoretically proposed LS metallic, HS insulating, and LS-CS insulating states for \RNO\ are characterized by entirely distinct band structures, which consequently give rise to completely different optical responses. Therefore, investigating the optical properties of \RNO\ and comparing them with first-principles calculations and the optical response of other nickelates may provide pivotal information on the nature of its ground state.

In this paper, we present an optical study of \NNO\ at 15 different temperatures between 5 and 300~K in the frequency range 30--50\,000~\icm\ (3.75~meV--6.25~eV). The optical conductivity of \NNO\ at 5~K consists of several different components: IR-active phonons, a far-infrared band $\alpha$, a mid-infrared band $\beta$, and a high-energy absorption edge. A comparison of our experimental results to theoretical calculations and the optical properties of other nickelates manifests that charge-stripe fluctuations exist in \NNO. The $\beta$ band and the high-energy absorption edge can be associated with electronic transitions between gapped Ni-$d_{x^2-y^2}$ states due to fluctuating charge stripes and the onset of electronic transitions between bands lying further away from $E_{F}$, respectively. In addition, the temperature-dependent optical spectral weight exhibits a $T^{2}$ law even for a cutoff frequency as high as 12\,000~\icm\ ($\sim$1.5~eV), which is likely to arise from strong correlation effects.

%%%%%%%%%%%%%%%%%%%%%%%%%%%%%%%%%%%%%%%%%%%%%%%%%%%%%%%%%%%%%%%%%%%%%%%%%%%%%%
%
% Experiment
%

%%%%%%%%%%%%%%%%%%%%%%%%%%%%%%%%%%
% Figure 1
\begin{figure}[tb]
\includegraphics[width=0.95\columnwidth]{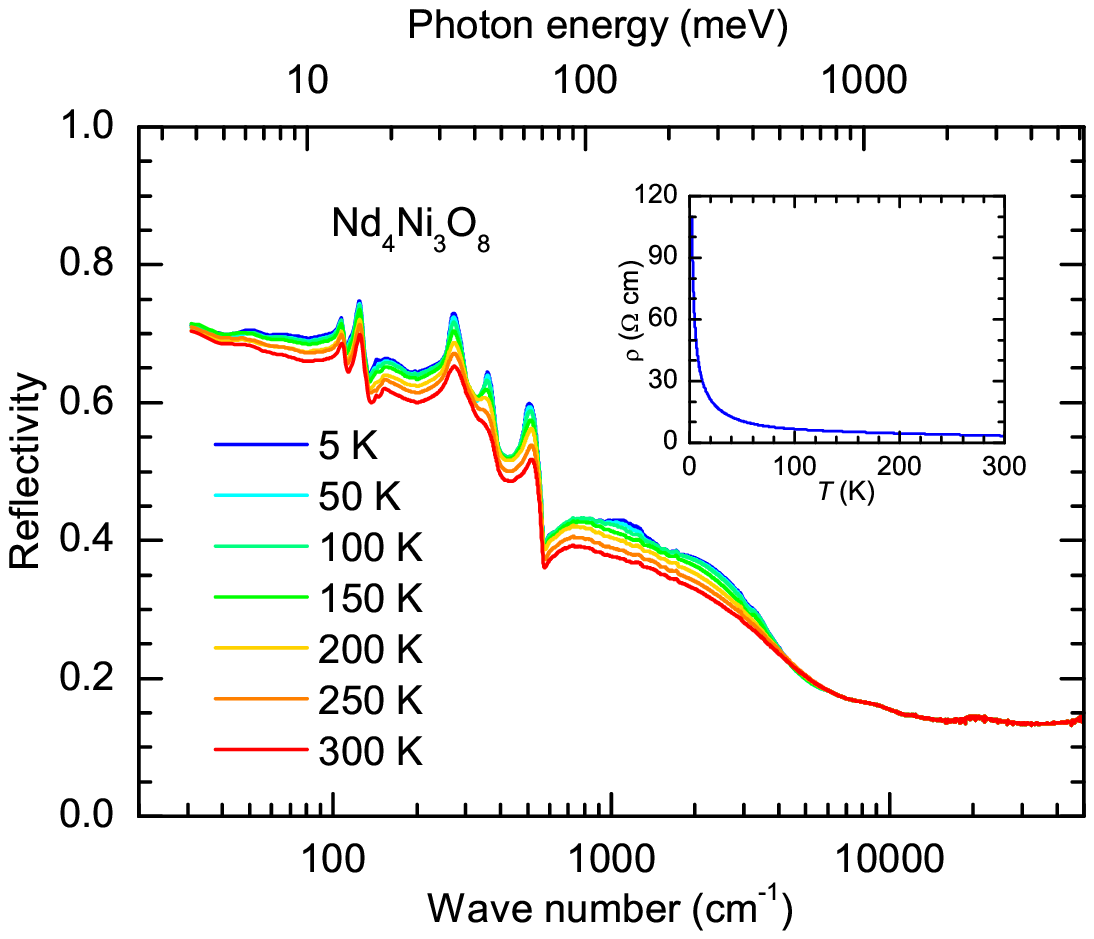}
\caption{Reflectivity of \NNO\ up to 50\,000~\icm\ at several selected temperatures from 300 down to 5~K. Inset: Resistivity of \NNO\ as a function of temperature.}
\label{NRef}
\end{figure}
High-purity polycrystalline samples of \NNO\ were obtained by reacting Nd$_{4}$Ni$_{3}$O$_{10}$ with CaH$_{2}$~\cite{Li2021SCPMA,Li2020CM}. The resistivity $\rho$ of our sample, as shown in the inset of Fig.~\ref{NRef}, rises with decreasing temperature, which is a clear signature of insulating behavior~\cite{Li2021SCPMA}. The near-normal incidence reflectivity $R(\omega)$ of \NNO\ was measured at 15 different temperatures between 5 and 300~K in the frequency range from 30 to 12\,000~\icm\ using a Bruker Vertex 80v Fourier transform spectrometer. An \emph{in situ} gold overfilling technique~\cite{Homes1993} was utilized to obtain the absolute reflectivity. We then used an AvaSpec-2048$\times$14 fiber optical spectrometer to extend $R(\omega)$ to 50\,000~\icm\ at room temperature.

First of all, we would like to point out that although single crystalline samples of \LNO\ and \PNO\ have been synthesized using the high-pressure floating zone technique~\cite{Zhang2016PNAS,Zhang2017NP,Zhang2021arXiv}, the growth of single crystalline \NNO\ requires much higher pressure that can not be reached thus far. As a consequence, experimental studies of \NNO\ are limited to polycrystalline samples~\cite{Li2021SCPMA}. Optical data collected on a polycrystal reflects the mixture of the optical responses from all three crystal axes, which may not be reliable for a quantitative analysis. Nevertheless, previous optical studies on polycrystals have shown that a qualitative analysis of optical data from polycrystalline samples, in combination with transport studies, can also provide useful information about the charge dynamics of the materials~\cite{Bonn1987PRL,Bonn1987PRB,Chen2008PRL,Dubroka2008PRL,Inaba1995PRB,Ueda2012PRL,Ueda2016PRB,Fujioka2021PRB}.

Figure~\ref{NRef} shows the measured $R(\omega)$ of \NNO\ up to 50\,000~\icm\ at several representative temperatures from 300 down to 5~K. The far-infrared $R(\omega)$ below about 600~\icm\ is dominated by IR-active phonons (sharp features). Below the phonon bands, $R(\omega)$ approaches a constant that is much lower than unity, which is the prototypical optical response of an insulator~\cite{Jiang2020PRB,Reijnders2014PRB,Lobo2007PRB}. This is consistent with the transport measurement, as shown in the inset of Fig.~\ref{NRef}, which also reveals insulating behavior in the temperature range of 2--300~K~\cite{Li2021SCPMA}. Above the IR-active phonon bands, $R(\omega)$ exhibits a broad hump-like feature in the frequency range 600--5000~\icm; a similar feature has been reported in the charge-stripe-ordered La$_{2-x}$Sr$_{x}$NiO$_{4}$~\cite{Ido1991PRB,Bi1993PRB} and La$_{2}$NiO$_{4+\delta}$~\cite{Homes2003PRB}.

%%%%%%%%%%%%%%%%%%%%%%%%%%%%%%%%%%%%%%%%%%%%%%%%%%%%%%%%%%%%%%%%%%%%%%%%%%%%%%
%
% Data analysis
%

%%%%%%%%%%%%%%%%%%%%
% Figure 2
\begin{figure}[tb]
\includegraphics[width=0.95\columnwidth]{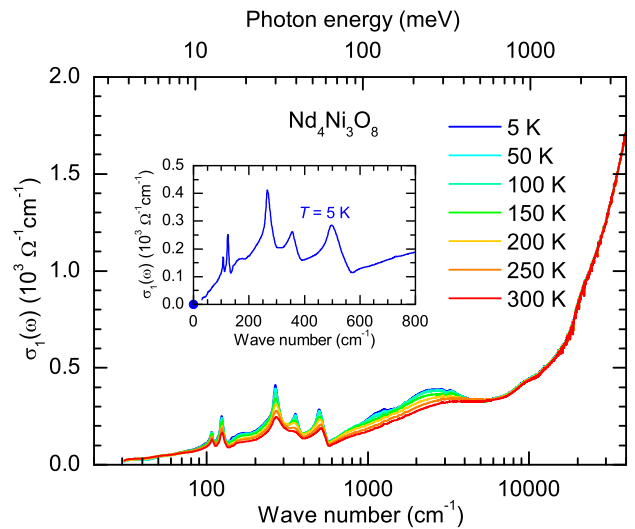}
\caption{The real part of the optical conductivity $\sigma_{1}(\omega)$ of \NNO\ at different temperatures from 300 down to 5~K. The inset shows $\sigma_{1}(\omega)$ in the far-infrared range at $T = 5$~K. The blue solid circle at zero frequency denotes the dc conductivity from transport measurements.}
\label{NS1}
\end{figure}
In order to obtain further information about the charge dynamics of \NNO, the real part of the optical conductivity $\sigma_{1}(\omega)$ is determined through a Kramers-Kronig analysis of $R(\omega)$. Since $R(\omega)$ exhibits a tendency towards saturation in the far-infrared range, a constant was used for the low-frequency extrapolation. On the high-frequency side, we adopted a constant $R(\omega)$ up to 12.5~eV followed by a free-electron ($\omega^{-4}$) response. Figure~\ref{NS1} displays $\sigma_{1}(\omega)$ of \NNO\ at different temperatures from 300 down to 5~K. The inset shows $\sigma_{1}(\omega)$ at 5~K in the far-infrared range, and the blue solid circle at zero frequency denotes the dc conductivity $\sigma_{\text{dc}}$ at 5~K from transport measurements. The zero-frequency extrapolation of the optical conductivity $\sigma_{1}(\omega\rightarrow0)$ agrees quite well with $\sigma_{\text{dc}}$, testifying to the self-consistency of our experimental results. The far-infrared $\sigma_{1}(\omega)$ is dominated by IR-active phonon modes which are characterized as sharp peaks below 600~\icm. The mid-infrared $\sigma_{1}(\omega)$ features a broad absorption band lying in the frequency range of 600--5000~\icm, which grows in amplitude as the temperature is lowered. Above 10\,000~\icm, $\sigma_{1}(\omega)$ increases sharply, leading to an absorption edge at about 20\,000~\icm.

\begin{figure}[tb]
\includegraphics[width=0.95\columnwidth]{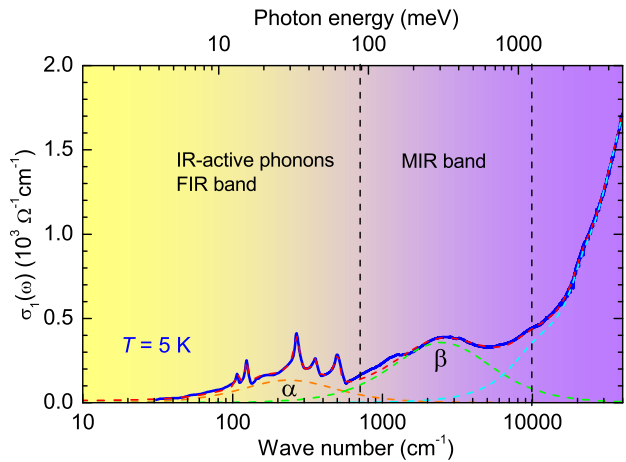}
\caption{The blue solid curve denotes $\sigma_{1}(\omega)$ of \NNO\ measured at 5~K. The red dashed line through the data is the fitting result, which is decomposed into an FIR band $\alpha$ (orange dashed line), an MIR component $\beta$ (green dashed line), and a high-energy absorption edge (cyan dashed line).}
\label{NFit}
\end{figure}

The measured $\sigma_{1}(\omega)$ of \NNO\ can be fit to a series of Lorentzian oscillators,
%%%%%%%%%%%%%
%
% Eq.1
%
\begin{equation}
\sigma_{1}(\omega) = \frac{2\pi}{Z_{0}} \sum_{j} \frac{\gamma_{j} \omega^{2} \Omega_{j}^{2}}{(\omega_{j}^{2} - \omega^{2})^{2} + \gamma_{j}^{2} \omega^{2}},
\label{Lorentz}
\end{equation}
where $Z_{0} \simeq 377~\Omega$ is the vacuum impedance; $\omega_{j}$, $\gamma_{j}$ and $\Omega_{j}$ correspond to the resonance frequency, linewidth and strength of the $j$th oscillator, respectively. The red dashed line in Fig.~\ref{NFit} represents the fitting result at 5~K, which reproduces the measured $\sigma_{1}(\omega)$ (blue solid line) quite well. As illustrated in Fig.~\ref{NFit}, the fit allows us to decompose $\sigma_{1}(\omega)$ of \NNO\ at 5~K into several different components: (i) IR-active phonon modes which are not shown in the figure; (ii) a far-infrared (FIR) absorption band $\alpha$ under the phonon modes (orange dashed line); (iii) a mid-infrared (MIR) absorption band $\beta$ centered at about 2500~\icm\ (green dashed line); (iv) an absorption edge at about 20\,000~\icm\ (cyan dashed line). In the following, we compare our experimental results to first-principles calculations and the optical response of other nickelates, aiming at understanding the origins of these components (except for the phonon modes) in $\sigma_{1}(\omega)$ and the ground state of \NNO.

%%%%%%%%%%%%%%%%%%%%
% Figure 3

First of all, it should be noted that a Drude component is absent in $\sigma_{1}(\omega)$ of \NNO, agreeing well with the insulating behavior revealed by transport measurements~\cite{Li2021SCPMA}. Nevertheless, we would like to mention that in systems with strong electronic anisotropy, such as cuprates and iron pnictides, the optical response of a polycrystalline sample is dominated by the weakly conducting $c$ axis~\cite{Bonn1987PRB,Bonn1987PRL,Homes1993PRL,Chen2008PRL,Dubroka2008PRL,Xu2020PRB}, while the Drude response associated with the metallic $ab$ plane is significantly suppressed. Hence, the absence of a Drude component in $\sigma_{1}(\omega)$ of the polycrystalline \NNO\ points towards two possibilities: either this material is insulating in both the $ab$ plane and the $c$ direction, or it is highly anisotropic with a metallic $ab$ plane and an insulating $c$ direction. The latter possibility may be ruled out by the insulating transport properties of the polycrystalline \NNO~\cite{Li2021SCPMA}, because for compounds that are metallic in the $ab$ plane but insulating along the $c$ axis, for example cuprates and iron pnictides, $\rho(T)$ of a polycrystalline sample has been found to exhibit prominent metallic behavior~\cite{Dubroka2008PRL,Takagi1992PRL,Wu2017PRM}.

%%%%%%%%%%%%%%%%%%%%
% Figure 4
\begin{figure}[tb]
\includegraphics[width=\columnwidth]{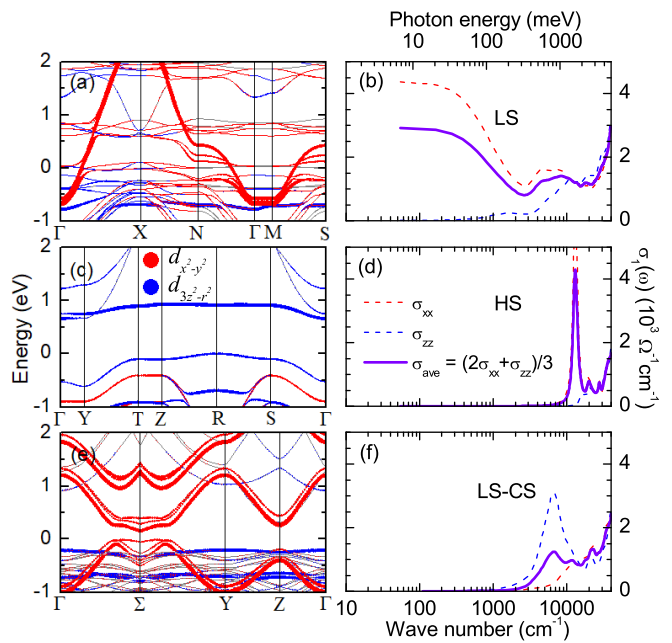}
\caption{(a), (c) and (e) display the calculated band structures of \NNO\ for the LS metallic, HS insulating and LS-CS insulating states, respectively. (b), (d) and (f) depict the calculated optical conductivity spectra based on the band structures in (a), (c) and (e), respectively.}
\label{NS1Cal}
\end{figure}

To gain more insight into the origins of the different components in $\sigma_{1}(\omega)$, we calculated the band structure of \NNO\ and corresponding $\sigma_{1}(\omega)$ using density functional theory (DFT)~\cite{Schwarz2002CPC,Schwarz2003CMS,Abt1994PB,Ambrosch-Draxl2006CPC}. The details about the calculations are given in the appendix. Since $\sigma_{1}(\omega)$ of a polycrystal consists of contributions from both the $ab$ plane and the $c$ direction, we calculated both the in-plane $\sigma_{xx}(\omega) = \sigma_{yy}(\omega)$ and out-of-plane $\sigma_{zz}(\omega)$ optical conductivities, and then compare the average $\sigma_{\text{ave}} = (\sigma_{xx}+\sigma_{yy}+\sigma_{zz})/3 = (2\sigma_{xx}+\sigma_{zz})/3$ to the measured $\sigma_{1}(\omega)$ of \NNO.

Figure~\ref{NS1Cal}(a) displays the calculated band structure of the LS state of \NNO\ which reveals a metallic phase with the partially filled Ni-$d_{x^2-y^2}$ bands crossing the Fermi level, in good agreement with previous calculations for La$_{4}$Ni$_{3}$O$_{8}$~\cite{Poltavets2010PRL,Sarkar2011PRB,Pardo2012PRB} and Pr$_{4}$Ni$_{3}$O$_{8}$~\cite{Botana2017PRM,Karp2020arXiv}. The calculated $\sigma_{xx}$ and $\sigma_{zz}$ for the LS state are shown as red and blue dashed lines respectively in Fig.~\ref{NS1Cal}(b). A pronounced Drude component, the optical signature of metallic behavior, can be seen in $\sigma_{xx}$. Although the absence of Drude response in $\sigma_{zz}$ implies insulating behavior along the $c$ direction, the metallic $\sigma_{xx}$ would inevitably produce coherent electronic transport in the temperature-dependent $\rho(T)$ of the polycrystalline \NNO. The robust insulating behavior in $\rho(T)$, as shown in the inset of Fig.~\ref{NRef} and Ref.~\cite{Li2021SCPMA}, suggests that the LS metallic state fails in describing the experimental results of \NNO. Thus, models yielding an insulating ground state should be invoked.

We recall that \LNO\ is also highly insulating at low temperatures. In order to account for the insulating behavior in \LNO, Pardo and Pickett consider the Ni trilayer as a molecular trimer with in-plane AFM order, from which on-site repulsion $U$ gives rise to a Mott insulating state, i.e. the HS insulating state~\cite{Pardo2010PRL}. Alternatively, Botana et al. have pointed out that an insulating ground state, the LS-CS insulating state, can also be obtained from a combination of charge-order-related structural distortions and magnetic order~\cite{Botana2016PRB}. We apply these two approaches to \NNO\ and derive $\sigma_{xx}$, $\sigma_{zz}$, as well as $\sigma_{\text{ave}}$ from the corresponding electronic structure in each case. The details about the theoretical calculations can be found in the appendix.

The calculated band structure of \NNO\ for the HS insulating state, which is similar to the previous calculation for \LNO~\cite{Pardo2010PRL}, is traced out in Fig.~\ref{NS1Cal}(c). A large (direct) gap of $\sim$1--1.5~eV opens between the $d_{3z^2-r^2}$ states that are depicted as blue solid circles. Figure~\ref{NS1Cal}(d) shows the optical conductivity spectra calculated from the band structure in Fig.~\ref{NS1Cal}(c). A sharp peak arising from direct interband transitions between the $d_{3z^2-r^2}$ bands can be identified at about 13\,000~\icm\ in $\sigma_{xx}$. On the low-frequency side, $\sigma_{xx}$ and $\sigma_{zz}$ vanish below 10\,000 and 15\,000~\icm, respectively. It is immediately obvious that the measured $\sigma_{1}(\omega)$ of \NNO\ contrasts with the calculated $\sigma_{\text{ave}}$ [purple solid line in Fig.~\ref{NS1Cal}(d)]: (i) absorption bands, i.e. $\alpha$ and $\beta$, exist in the low-frequency (below $\sim$10\,000~\icm) region of the measured $\sigma_{1}(\omega)$, while $\sigma_{\text{ave}}$ is vanishingly small in the same spectral range; (ii) the sharp peak at 13\,000~\icm\ in the calculated $\sigma_{\text{ave}}$ is not observed in the measured $\sigma_{1}(\omega)$. Such considerable discrepancies between the calculated $\sigma_{\text{ave}}$ and the measured $\sigma_{1}(\omega)$ do not favor the HS insulating state in \NNO.

Next, we consider the LS-CS insulating state. Figure~\ref{NS1Cal}(e) shows the calculated band structure of \NNO\ for the LS-CS insulating state, which also resembles the previous result in \LNO~\cite{Botana2016PRB}. A much smaller gap is formed between the Ni $d_{x^2-y^2}$ bands (red solid circles), which is completely different from the HS insulating state. The calculated $\sigma_{xx}$ and $\sigma_{zz}$ are shown in Fig.~\ref{NS1Cal}(f) as red and blue dashed lines, respectively. While $\sigma_{xx}$ diminishes with decreasing frequency and disappears at $\sim$3000~\icm, $\sigma_{zz}$ exhibits a broad absorption peak at $\sim$6000~\icm\ with a tail extending down to $\sim$1000~\icm. The peak and the low-energy tail in $\sigma_{zz}$ are predominantly contributed by interband electronic transitions across the gap between the Ni $d_{x^2-y^2}$ bands. Intriguingly, the calculated $\sigma_{\text{ave}}$ qualitatively reproduces the measured $\sigma_{1}(\omega)$ in the MIR and high-frequency range, hinting that the ground state of \NNO\ might be (or is in close proximity to) an LS-CS insulator as previously proposed for \LNO~\cite{Botana2016PRB}. Within this framework, the $\beta$ band may be ascribed to electronic transitions between the gapped $d_{x^2-y^2}$ bands, and the high-frequency absorption edge is associated with the onset of electronic transitions involving bands lying further away from $E_{F}$.

It is noteworthy that the optical response of \NNO\ bears a remarkable resemblance to that of charge-stripe-ordered La$_{2-x}$Sr$_{x}$NiO$_{4}$~\cite{Ido1991PRB,Bi1993PRB,Katsufuji1996PRB,Calvani1996PRB} and La$_{2}$NiO$_{4+\delta}$~\cite{Homes2003PRB}. An MIR peak akin to the $\beta$ band in \NNO\ has also been observed in $\sigma_{1}(\omega)$ of these materials~\cite{Homes2003PRB,Ido1991PRB,Bi1993PRB,Katsufuji1996PRB,Calvani1996PRB}. While early studies have assigned the MIR peak to the formation of small polarons~\cite{Bi1993PRB,Katsufuji1996PRB,Calvani1996PRB,Jung2001PRB}, Homes et al.~\cite{Homes2003PRB} have taken into account the fact that the holes doped into the NiO$_{2}$ planes form a stripe order which has been established by neutron and X-ray scattering studies~\cite{Tranquada1994PRL,Yoshizawa2000PRB}, and consequently ascribed the MIR peak to transitions from filled valence states to empty mid-gap states associated with the charge stripes. Furthermore, as the MIR peak persists to temperatures far above the charge-stripe-ordering transition temperature $T_{\text{CO}}$~\cite{Katsufuji1996PRB,Homes2003PRB}, recent observations of short-range-ordered or dynamic charge stripes on the same temperature scale have naturally related it to the presence of charge-stripe fluctuations~\cite{Abeykoon2013PRL,Coslovich2013NC,Anissimova2014NC}. Given the striking resemblance between the optical response of \NNO\ and that of La$_{2}$NiO$_{4+\delta}$ and La$_{2-x}$Sr$_{x}$NiO$_{4}$, it is plausible to associate the $\beta$ band in \NNO\ with the emergence of charge stripes or charge-stripe fluctuations. Although the formation of charge stripes below the transition at 105~K in \LNO\ has been confirmed by the synchrotron X-ray diffraction~\cite{Zhang2016PNAS}, no phase transition has been detected in \NNO\ from room temperature down to 2~K~\cite{Li2021SCPMA}. In this case, the $\beta$ band in \NNO\ is most likely to signify charge-stripe fluctuations, i.e. short-range-ordered or dynamic charge stripes.

The presence of charge-stripe fluctuations or even a static charge-stripe order in \NNO\ is not surprising, because it is structurally and electronically identical to \LNO\ which has been found to order in charge stripes below the transition at 105~K~\cite{Zhang2016PNAS}. Although the relatively smaller atomic radius of Nd may suppress the static long-range order in \NNO, charge-stripe correlations should not be significantly affected by such a small difference. In addition, similar to the case of La$_{4}$Ni$_{3}$O$_{8}$~\cite{Botana2016PRB}, the calculated LS-CS state has a lower free energy than the HS molecular insulating state, representing a favorable ground state for \NNO.

We further notice that neither the HS nor the LS-CS insulating state can account for the $\alpha$ band in the measured $\sigma_{1}(\omega)$ of \NNO. A similar FIR absorption band has been observed in doped semiconductors Si:P and assigned to optical transitions between impurity states and the conduction band~\cite{Gaymann1995PRB}; Some bismuth-based topological insulators, e.g. Bi$_{2}$Te$_{2}$Se and Bi$_{2-x}$Ca$_{x}$Se$_{3}$, also exhibit such an impurity-related absorption band centered at about 200~\icm~\cite{Akrap2012PRB,Pietro2012PRB}. By analogy with Si:P and bismuth-based topological insulators, the $\alpha$ band in \NNO\ may be attributed to transitions involving in-gap impurity states. Alternatively, the $\alpha$ band may be an artifact arising from the polycrystalline nature of our sample. In La$_{2}$NiO$_{4+\delta}$ and La$_{2-x}$Sr$_{x}$NiO$_{4}$, the $ab$-plane $\sigma_{1}(\omega)$ exhibits a residual Drude response in the charge-stripe-fluctuation regime~\cite{Homes2003PRB,Coslovich2013NC}. Such a residual Drude response may also exist in the $ab$-plane $\sigma_{1}(\omega)$ of \NNO\ due to the fluctuations of charge stipes. In this case, the mixture of a weak metallic spectrum from the $ab$ plane and an insulating one from the $c$ direction may create an artifact structure, i.e. the $\alpha$ band in $\sigma_{1}(\omega)$. Another possible origin of the $\alpha$ band is the electronic excitations within the stripes. If the stripe order is neither static nor of long range, then low-energy excitations within the stripes may be present, giving rise to the $\alpha$ band in the measured $\sigma_{1}(\omega)$ of \NNO.

%%%%%%%%%%%%%%%%%%%%
% Figure 5
\begin{figure}[tb]
\includegraphics[width=0.9\columnwidth]{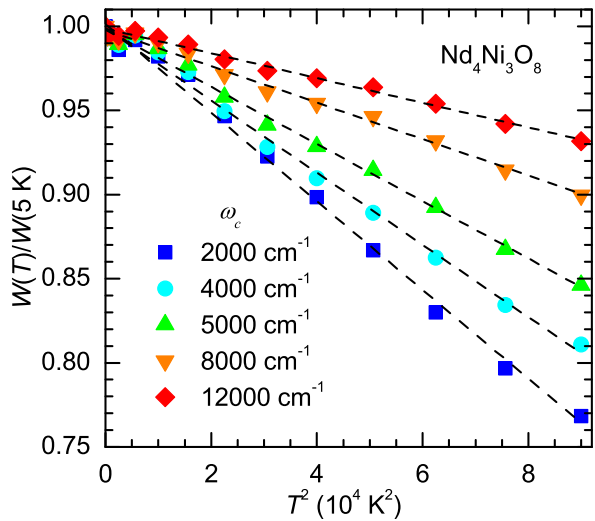}
\caption{Spectral weight up to different cutoff frequencies as a function of $T^2$. The dashed lines are linear fits.}
\label{NRSW}
\end{figure}
Finally, we examine the evolution of $\sigma_{1}(\omega)$ with temperature. It is noticeable in Fig.~\ref{NS1} that $\sigma_{1}(\omega)$ in the FIR and MIR region grows as the temperature is lowered. This is distinct from materials with charge stripes or charge-stripe fluctuations, such as La$_{2-x}$Sr$_{x}$NiO$_{4}$ and La$_{2}$NiO$_{4+\delta}$, in which a suppression of the low-frequency $\sigma_{1}(\omega)$ alongside an enhancement of $\sigma_{1}(\omega)$ in the MIR peak region has been observed~\cite{Homes2003PRB,Coslovich2013NC}. Hence, the unusual temperature dependence of the optical response in \NNO\ is likely to be dominated by a different effect.

The detailed temperature dependence of $\sigma_{1}(\omega)$ can be tracked by inspecting the optical spectral weight $W$ defined as
%%%%%%%%%%%%%
%
% Eq.2
%
\begin{equation}
W=\int_{0}^{\omega_c}\sigma_{1}(\omega)d\omega
\label{SW},
\end{equation}
where $\omega_{c}$ represents a cutoff frequency. The symbols in Fig.~\ref{NRSW} denote $W$ as a function of $T^{2}$ for different cutoff frequencies, and the dashed lines through the symbols are linear fits. $W$ increases upon cooling even for $\omega_{c}$ = 12\,000~\icm\ ($\sim$1.5~eV), indicating a spectral weight transfer from an energy range higher than $\sim$1.5~eV to the FIR-MIR region. A more intriguing observation here is that $W$ varies linearly with $T^{2}$ for all $\omega_{c}$'s we have chosen. Such behavior has been widely observed and extensively studied in cuprates~\cite{Molegraaf2002Science,Ortolani2005PRL,Toschi2005PRL,Carbone2006PRB,Carbone2006PRB1}. In a normal metal, e.g. gold~\cite{Ortolani2005PRL}, $W$ is linear in $T^{2}$ for $\omega_{c} < \omega_{p}$, but is temperature independent for $\omega_{c} \geq \omega_{p}$. Here, $\omega_{p}$ is the plasma frequency ($\omega_{p}^{2}$ is proportional to the carrier density). However, in cuprates~\cite{Molegraaf2002Science,Ortolani2005PRL,Toschi2005PRL,Carbone2006PRB,Carbone2006PRB1}, $W$ exhibits conspicuous linear dependence on $T^{2}$ even for $\omega_{c} \geq \omega_{p}$, which is in sharp contrast to a conventional metal. Dynamical mean-field theory calculations based on a strongly correlated Hubbard model have demonstrated that the $T^{2}$ law of $W$ in cuprates arises from strong correlation effects~\cite{Toschi2005PRL}. The phenomenon that $W$ varies linearly with $T^{2}$ in \NNO\ is peculiar, since it is an insulator without any observable free-carrier response. Although detailed temperature dependence of the optical spectral weight in insulating cuprates has not yet been reported, we notice that in the slightly doped but still insulating Bi$_{2}$Sr$_{2-x}$La$_{x}$CuO$_{6}$ (see Figs.~1(c) and 1(d) in Ref.~\cite{Lupi2009PRL}), as the temperature is lowered, while the spectral weight below about 80~\icm\ is suppressed, a much more pronounced increase of the spectral weight occurs between 80 and 2000~\icm, resulting in a net spectral weight gain in the FIR-MIR range. Such an FIR-MIR spectral weight increase at low temperatures can also be identified in the optical conductivity of the slightly doped insulating YBa$_{2}$Cu$_{3}$O$_{y}$ (Fig.~3 for $y = 6.28$ in Ref.~\cite{Lee2005PRB}). Considering these facts, the linear evolution of $W$ with $T^{2}$ in \NNO\ may be governed by the same mechanism as that in cuprates, i.e. strong correlation effects. Therefore, strong electronic correlation may also play an important role in \NNO.

%%%%%%%%%%%%%%%%%%
% Discussion
%

%%%%%%%%%%%%%%%%%%%%%%%%%%%%%%%%%%%%%%%%%%%%%%%%%%%%%%%%%%%%%%%%%%%%%%%%%%%%%%
%
% Conclusions
%
To summarize, we experimentally obtained the optical conductivity of \NNO\ at 15 different temperatures between 5 and 300~K in the frequency range 30--50\,000~\icm\ (3.75~meV--6.25~eV). Data fitting based on the Lorentz model allows us to decompose the optical conductivity at 5~K into IR-active phonons, an FIR $\alpha$ band, an MIR $\beta$ band, as well as a high-energy absorption edge. A comparison of the measured optical conductivity to theoretical calculations and the optical response of other nickelates suggests that \NNO\ features prominent charge-stripe fluctuations. The $\alpha$ band may possibly be associated with impurities, the mixture of an in-plane residual Drude response and our-of-plane insulating behavior, or electronic excitations within the stripes; the $\beta$ band and the high-frequency absorption edge can be ascribed to electronic transitions across the charge-stripe-fluctuation-induced gap between the Ni-$d_{x^2-y^2}$ states and the onset of transitions between other high-energy bands, respectively. The optical spectral weight varies linearly with $T^{2}$ even for a cutoff frequency as high as 12\,000~\icm\ ($\sim$1.5~eV). Such a $T^{2}$ law, which is also widely observed in cuprates, may be related to strong correlation effects in \NNO.

%%%%%%%%%%%%%%%%%%%%%%%%%%%%%%%%%%%%%%%%%%%%%%%%%%%%%%%%%%%%%%%%%%%%%%%%%%%%%%
%
% Acknowledgment
%

\begin{acknowledgments}
We thank Christopher C. Homes, Ricardo Lobo, Jiawei Mei, Yilin Wang, and Shunli Yu for illuminating discussions. We gratefully acknowledge financial support from the National Key R\&D Program of China (Grant No. 2016YFA0300401), the National Natural Science Foundation of China (Grants No. A2008/11874206, E0209/52072170, 12061131001), the Strategic Priority Research Program of Chinese Academy of Sciences (Grant No. XDB25000000), the Fundamental Research Funds for the Central Universities with Grant No. 020414380095, and Jiangsu shuangchuang program.

Jiahao Hao, Xinwei Fan and Qing Li contributed equally to this work.
\end{acknowledgments}

%%%%%%%%%%%%%%%%%%%%%%%%%%%%%
%
%   Appendix
%

\appendix

\section*{Appendix: theoretical calculations}
The band structure of \NNO\ and corresponding $\sigma_{1}(\omega)$ were calculated using density functional theory (DFT) implemented in the full-potential linearized augmented plane wave code WIEN2k~\cite{Schwarz2002CPC,Schwarz2003CMS,Abt1994PB,Ambrosch-Draxl2006CPC}. The structural information of \NNO\ was taken from Ref~\cite{Li2021SCPMA}. The Perdew-Burke-Ernzerhof (PBE) generalized gradient approximation (GGA)~\cite{Perdew1996PRL} was chosen as the exchange-correlation potential. To avoid the ambiguity of the 4$f$ electrons of Nd around the Fermi energy, we employed the so-called GGA$+U$ scheme with an effective Hubbard $U_{\text{eff}}$ = 9~eV on the 4$f$ electrons of Nd to repel them away from the Fermi level. This method has been used in previous calculations for Nd$_{1-x}$Sr$_{x}$NiO$_{2}$~\cite{Choi2020PRB,Ryee2020PRB} and \NNO~\cite{Fan2020JPCM}. We also applied $U_{\text{eff}}$ = 5~eV on the 3$d$ electrons of Ni, since the correlation effect between 3$d$ electrons of transition-metal atoms should be profound. Here, the value of $U_{\text{eff}}$ we chose for the 3$d$ electrons of Ni is close to those used in previous work on similar materials such as \LNO~\cite{Poltavets2010PRL,Pardo2010PRL,Botana2016PRB}, \NNO~\cite{Fan2020JPCM}, and Nd$_{1-x}$Sr$_{x}$NiO$_{2}$~\cite{Choi2020PRB,Ryee2020PRB}.

\begin{figure}[tb]
\includegraphics[width=\columnwidth]{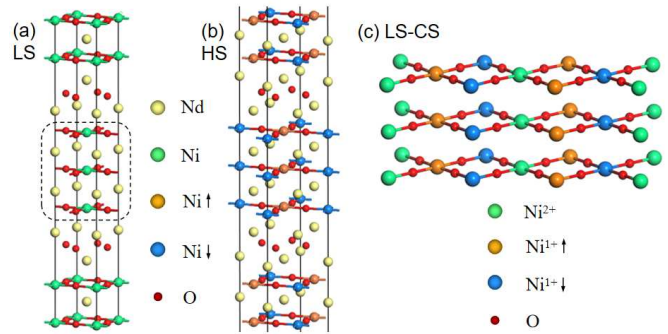}
\caption{(a) The crystal structure of \NNO. The dashed frame highlights the NiO$_{2}$ trilayer. (b) The $\sqrt{2}a \times \sqrt{2}a \times c$ supercell with an in-plane AFM order, which is used for the calculations of the HS insulating state. (c) The charge ordering pattern we adopt for the calculations of the LS-CS insulating state.}
\label{NLattice}
\end{figure}

Based on the crystal structure of \NNO\ shown in Fig.~\ref{NLattice}(a), our theoretical calculations produce an LS metallic phase represented by the band structure in Fig.~\ref{NS1Cal}(a). For this LS metallic state, no matter whether $U_{\text{eff}}$ is applied, \NNO\ always remains metallic. The same behavior has been found in \LNO~\cite{Poltavets2010PRL}.

For the HS insulating state, we adopt the approach that was initially proposed by Pardo and Pickett to account for the insulating behavior in \LNO~\cite{Pardo2010PRL}. Specifically, we consider a $\sqrt{2}a \times \sqrt{2}a \times c$ supercell with an in-plane AFM order, as depicted in Fig.~\ref{NLattice}(b) where the orange and blue spheres denote Ni with different spin directions. With this supercell, the calculated band structure of \NNO, as shown in Fig.~\ref{NS1Cal}(c), exhibits a large gap between the Ni $d_{3z^2-r^2}$ bands, closely resembling the previous calculation for \LNO~\cite{Pardo2010PRL}. A larger $U_{\text{eff}}$ for the Ni 3$d$ electrons increases the gap, but makes no essential difference in the electronic structure. However, a minimum $U_{\text{eff}} \simeq 3$~eV is required to open a gap for the HS insulating state.

%%%%%%%%%%%%%%%%%%%%
% Figure 6

The LS-CS insulating state, which was obtained by Botana et al.~\cite{Botana2016PRB} as the ground state of \LNO, requires a $3\sqrt{2}a \times \sqrt{2}a \times c$ supercell with the charge/spin pattern illustrated in Fig.~\ref{NLattice}(c). The green spheres correspond to non-magnetic Ni$^{2+}$ ions; the orange and blue spheres represent Ni$^{1+}$ ions with different spin directions. Using such a supercell, we calculated the band structure of \NNO\ for the LS-CS state [Fig.~\ref{NS1Cal}(e)], which reveals a small gap between the Ni $d_{x^2-y^2}$ states, similar to the previous calculation for \LNO~\cite{Botana2016PRB}. While increasing $U_{\text{eff}}$ for the Ni 3$d$ electrons leads to an expansion of the gap, no substantial change in the band structure is observed. In addition, a gap of 0.12~eV still remains even without introducing a Coulomb $U$ on the Ni 3$d$ electrons for the LS-CS state. These effects have also been noticed in previous work on \LNO~\cite{Botana2016PRB}.

Finally, we would like to remark that the LS metallic state is much more unstable than the HS insulating state, in agreement with the calculations for \LNO~\cite{Liu2012JPCM}, while the LS-CS insulating state is about 0.4~eV/Ni more stable than the HS insulating state, which is also consistent with previous calculations for \LNO~\cite{Botana2016PRB}.

%%%%%%%%%%%%%%%%%%%%%%%%%%%%%%%%%%%%%%%%%%%%%%%%%%%%%%%%%%%%%%%%%%%%%%%%%%%%%%%
% The bibliography (BibTeX)
%

\end{document}